\documentclass[a4paper,11pt]{article}
\usepackage{pos}

\title{Is flavor discrete?}

\author*[a,b]{Alexei Y Smirnov}

\affiliation[a]{Max-Planck-Institute f\"{u}r Kernphysik, Saupfercheckweg 1, Heidelberg, Germany}
\affiliation[b]{International Centre for Theoretical Physics, 
Strada Costiera 11, Trieste, Italy}

\emailAdd{smirnov@ictp.it}

\abstract{
Relevance  of the  discrete symmetries
for  explanation of the observed flavor structures in the leptonic sector is considered. 
Achievements of the "traditional'' discrete symmetry approach 
and the modular symmetry approache are confronted.
Minimal models with small number of parameters do not work.
Complication of  symmetry prescriptions allow to introduce
new free parameters and thus describe the data but simultaneously bring two connected problems:
(i) problem of "missing" representations, and (ii) problem of selection of certain point 
in huge discrete parameter space formed by possible charge assignments. 
Both problems must be addressed in complete model. Alternatively, one can keep 
minimal symmetry prescription but extend models introducing 
new physics unrelated to the original flavor symmetry. 
The low energy predictions can be "polluted" by additional physics such as effects
of coupling to hidden sector, RGE running, decoupling of heavy degrees of freedom, 
mixing with sterile neutrinos, {\it etc.} The set up with discrete symmetries in the Hidden sector
which communicates to  the visible one via the neutrino portal 
and the basis fixing symmetry still looks promising. 
}

\FullConference{%
9th Symposium on Prospects in the Physics of Discrete Symmetries (DISCRETE2024)
December 2-6,  2024\\
Ljubljana Slovenia}

\begin{document}
\maketitle

\section{Introduction} 

The question in the title  is essentially about relevance of the discrete symmetries 
for explanation of the  observed flavor structures (see reviews 
\cite{Feruglio:2019ybq}, \cite{Ding:2023htn}, \cite{Chauhan:2023faf}, \cite{Xing:2020ijf} 
and references therein, also \cite{ff} and \cite{Trautner:2023xai}).  
They include quark masses and mixing, lepton masses and mixing,  
CP violating phases. Are discrete symmetries behind these flavor structures?  
If so, the  doubt is  that we apply them  in right way. 
Probably something is missing that eventually will turn out to be the key for understanding 
whole the picture.   

The  talk covers the following issues (i) Bottom-up: data and hints;  
(ii) From traditional symmetry to modular symmetry. Two problems;  
(iii) On model building. Examples;
(iv) Further complications and ``pollution''.

\section{Bottom-up: data and hints} 

Recall, that the present phase of developments in the flavor physics  was triggered by discovery 
of the approximate Tri-Bimaximal (TBM)  
lepton mixing \cite{Harrison:2002er} and wide applications of  the non-Abelian discrete 
symmetries \cite{aa44}. This led to elaboration of the residual symmetry approach, developments of flavons dynamics, 
and more recently, to the modular symmetry approach.

The approximate TBM   can be presented as
\begin{eqnarray} 
U_{TBM}^{appr} &  = & U_{23}(\pi/4)U_{12} (\theta_{12}) +
\delta(U_{TBM}) 
\nonumber\\   
& = & \left(\begin{matrix}   
\sqrt{2/3}(1 + 0.010)  &   \sqrt{1/3}  (1 - 0.049)  &    0 \pm 0.149 \\ 
-  \sqrt{1/6}(1 - 0.105)   &  \sqrt{1/3}(1 + 0.130)   & - 
\sqrt{1/2}(1 - 0.076)\\  
-  \sqrt{1/6}(1 - 0.130)   &   \sqrt{1/3}(1 - 0.098)   &    
\sqrt{1/2}(1 + 0.024) 
\end{matrix} 
\right), 
\label{tbmmat}
\end{eqnarray}
where  $\sin^2\theta_{12} = 1/3$. For the observed values of matrix elements  
we used the central values from the NuFIT v6.0 
\cite{Esteban:2024eli}. 
According to (\ref{tbmmat}) the relative deviations of the matrix elements 
from TBM values range from 1$\%$ to 13 $\%$ and do not show clear 
systematics.  The TBM with the CKM ``dressing'': 
$U_{TBM}^{appr} = V_{CKM}^\dagger U_{TBM}$ (which is basis of of the Quark- Lepton Complementarity, QLC) 
requires much smaller corrections.  

Exact TBM provides clear signature of symmetry.  
In fact, the underlying symmetry was derived starting from the TBM form of mixing  \cite{Lam:2008rs} 
(see Fig.1):  
1)  mass matrices of charged leptons and neutrinos 
were constructed which lead to the TBM, 
2) transformations that leave invariant 
the matrices were found, 
3) covering groups for these transformations are identified.  
It turns out that  generators of these 
transformations coincide with the generators of  
$S_4$, or $A_4 \times Z_2$.  
There is no clear  relationship of mixing  with masses  (mass ratios). 

There are two key elements in realization of this symmetry building. 

{\it I.  Decoupling of lepton (in particular neutrino) masses from  mixing.} 

(i) Certain relations  between  elements
of the neutrino mass matrix in the flavor basis fix mixing. 
In particular, the equalities  $m_{\mu\mu} = m_{\tau\tau}$,   
$m_{e\mu} = - m_{e\tau}$,  $m_{ee} + m_{e\mu} 
= m_{\mu\mu} + m_{\mu\tau}$ produce the TBM mixing
independently of values of these elements. 

(ii) Combinations of the mass 
matrix elements fix the  values of neutrino masses: 
e.g. $m_3 = m_{\tau\tau} - m_{\mu\tau}$, $m_2 = m_{ee} + 2m_{e\mu}$. 

The decoupling of masses and mixing is always possible, but the key point 
is that the relations which  lead to required mixing should 
be simple enough to be consequences of symmetries.  \\

{\it II. Residual symmetry approach}

Mixing is a result of different ways of the flavor symmetry breaking 
in the  neutrino and charged lepton (Yukawa) sectors (see Fig.1). 
Different generators of original group remain unbroken in $\nu-$ and $l-$ sectors.
These different ways are usually related to different flavor charges 
of the right handed components of fermions. The latter challenges the  L-R symmetry and unification, 
unless further complications are introduced.

\begin{figure}
\begin{center}
\includegraphics[width=0.6\linewidth]{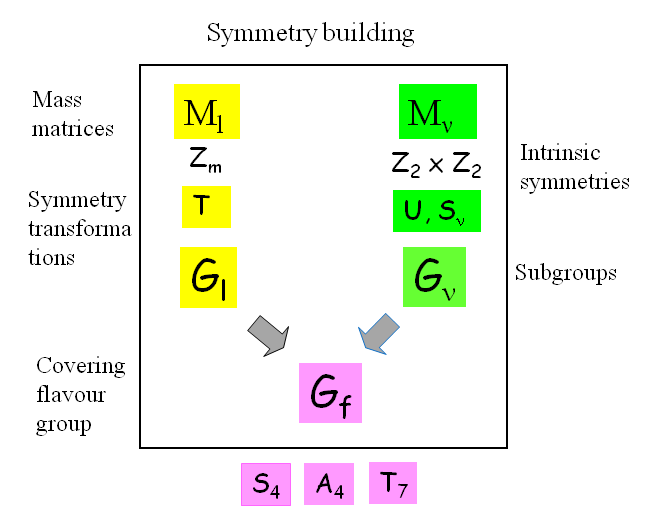}
\caption{
Residual symmetry scheme. Residual symmetries of 
the mass matrices of neutrino and charged leptons.
$T, U, S$ are generators  of symmetry transformations of mass matrices, 
$S_\nu M_\nu S_\nu^T  = M_\nu $, {\it etc}. }
\end{center}
\label{fig:resid}
\end{figure}

The Yukawa couplings are functions of flavon
fields, $\phi_{ki}$, which form $k-$multiplets 
of flavor group. 
Flavons - scalar fields,  singlets of SM symmetry group,   
break flavor symmetries. 
Set of independent Yukawa couplings 
$\{ h_i \}$ $i = 1, ..k$ is substituted as  
\begin{equation}
\{ h_i \}  \rightarrow \alpha \frac 
{\langle \phi_{ki} \rangle}{\Lambda}, 
\label{eq:yukawas}
\end{equation}  
and in general, 
\begin{equation}
\{ h \}  \rightarrow \alpha \frac{\Pi_{k = 1 - n} 
\langle \phi_k \rangle}{\Lambda^n}, 
\label{eq:yukawasg}
\end{equation}  
where  $i$ enumerates component of $k-$ plet,  $\alpha$ is free parameter and  $\Lambda$ 
is the scale of flavor physics.  
The couplings are given by VEV's of components of flavon multiplet, 
thus reducing the problem of  coupling fixing 
to determination of VEV's of flavons, that is,  to construction of  relevant flavon potential.

It turns out to be difficult to realize the approach and obtain TBM in consistent gauge theory.   
The problems are shifted to the flavon sector: to VEVs of flavons, VEV alignments, construction of certain 
flavons potential. There are many assumptions and many 
free parameters involved, furthermore, additional (auxiliary) 
fields should be introduced, {\it etc.} (see e.g. \cite{Smirnov:2011jv}).

At this point it is worthwhile to make one step back: do we understand and interpret 
the observations correctly? Is the approximate TBM accidental? 
There is no simple and exact relations between observables. 
Apart from clear qualitative flavor features: hierarchies of masses, large lepton mixing,  
it seems there is randomness in the data. No good description of data 
(and corresponding mass matrices) in terms of small number of parameters 
and assumptions is obtained. In other words, 
there is no simple parametrization of the mass matrices. 
The required mass matrix elements 
can be parametrized in a spirit of Froggatt-Nielsen mechanism 
\cite{Froggatt:1978nt} as 
\begin{equation}
m_{\alpha \beta}  = A_{\alpha \beta} 
\epsilon^{n_{\alpha \beta}} m_0,  \, \, \, \, \, \,       
\epsilon = \sin \theta_C.   
\label{eq:matelem}
\end{equation}
The prefactors $A_{\alpha \beta}$ being of the order 1 (0.7 - 1.5  at best) 
turn out to be  rather random for checked  choices of integers 
$n_{\alpha \beta}$.  
Randomness implies that there is no simple explanation of  data and simple theory behind. 
AI searchers of order and systematics can be done.   
Parameters may be traded with  assumptions. \\

\section{From traditional symmetry to modular symmetry. Two problems.}

Going from  the ``traditional'' symmetry approach to the modular symmetry approach we substitute 
the  Yukawa couplings as
\begin{equation}
\alpha \frac{\langle\phi_n\rangle}{\Lambda}          
\rightarrow 
\alpha Y_n^{(k_Y)} (\tau), 
\label{eq:tomodf}
\end{equation}
where $Y_n^{(k_Y)}$ are the modular forms  - the  known functions of moduli 
field $\tau$  that (i) depend on the weight $k_Y$ (parameters 
of symmetry transformations),  and (ii) transform under representation  
$n$ of the discrete symmetry group. In (\ref{eq:tomodf})
$\alpha$ are free coupling constants.   Certain weights are ascribed also to other 
fields which enter Yukawa interaction terms: fermions $k_f$ and Higgses $k_H$.   
Invariance of interactions with respect to the modular 
transformation requires 
\begin{equation}
\sum_f k_f  + k_H + k_Y = 0. 
\label{eq:conservation}
\end{equation}

Let us compare the modular and  traditional discrete symmetry applications. 
In the modular case the couplings 
$$
\left[Y_{n 1} (\tau), Y_{n 2} (\tau), ... , Y_{n n} (\tau) \right]   
$$
form multiplets of symmetry group similarly to flavon multiplets.
For a given symmetry level $N$  and weight $k$,  $Y$ are known 
functions of the  VEV of moduli $\langle \tau \rangle$ which is treated as free parameter.    
The key point is that $\tau$ is the same for all multiplets of couplings. 
In the traditional case the couplings are proportional to  
$$
\left[ \langle \phi_{n1}\rangle,  \langle \phi_{n2}\rangle , ... ,  \langle \phi_{n n}\rangle) \right]
$$
and the flavon VEVs depend on parameters  
of potential that are not controlled (directly) by symmetry (as well as $\tau$). 
The VEV's are independent for different representations. 
Thus, multiplet of the modular forms  $Y_b^{(k)} (\tau)$  acts as multiplet 
of flavons $\phi_b$ with specially selected VEVs:  
\begin{equation}
\left[ Y_{1n} (\tau), Y_{n2} (\tau) ... Y_{nn} (\tau)\right]  
= 
\frac{1}{\Lambda} \left[\langle \phi_{n1} \rangle, 
\langle \phi_{n 2} \rangle , ... , 
\langle \phi_{n n}\rangle \right]  
\label{eq: equa}
\end{equation}
and can do the same.   
Restriction from weights can be reproduced 
in the flavons approach by, e.g., additional $U(1)$ symmetry.   
In general, with flavons there is  more freedom to fit data.   
Modular symmetry models are expected to be  more restricted. 
However, substantial freedom exists (and can be introduced) 
also in the case of modular symmetries.  
General form of the Yukawa interaction term  (e.g., for quarks) is   
$$
\alpha  Y_b^{(k)} (\tau)  Q u^c H_u  
$$
with  $\alpha$ and $\tau^R , \tau^I$ used as free parameters.   
Additional freedom is in selection of weight  $k$ of $Y$.  
Invariance is then achieved by adjusting  weights of the fields 
(\ref{eq:conservation}).  
For high weight $k$,  $Y_b^{(k)}$ can form several multiplets of different  
dimension $b$: e.g. $Y_3^{(k)}$, $Y_2^{(k)}$.   
Correspondingly, several independent  invariant terms with free couplings can be introduced  
in the Lagrangian for the same fermions bilinear  $(Q u^c)$. 


More parameters and more freedom exist in  
models with non-minimal symmetries. 
In this case however two generic and actually related problems appear 
which were not appreciated much so far.   
For illustration of the problems we will use example from another flavor physics. \\

{\it I. Missing representations.}  Recall the story of hadron classification and  eightfold way.   
The dimension and quantum numbers of components of families of mesons and baryons show that they form 
representation 8 of the SU(3) symmetry group. The irreducible representation 8 
appears for mesons as $3 \times  3^* = 8 + 1$. Where are particles from the fundamental 
and other representations of SU(3)?
It turned out that a lot of new physics was uncovered answering this question. 

(i) The  fundamental rep. 3 is filled by quarks. 

(ii) $3 \times 3 = 6 + 3$: triplets and 6-plets  of  
hadrons are not observed because of colour,  QCD, and confinement.  

(iii) $3 \times 3 \times 3 =  1 + 8 + 8 + 10$.    
Both octets and decuplets of baryons were discovered. Moreover, 
the $10$th component of $10$ plet was predicted and 
then discovered which was confirmation  of the theory. 

(iv) Hadrons in higher representations - tetraquarks and  pentaquarks  were also  observed. 

The lesson is that Nature does not ``like'' to miss representation  
and fills in them starting from lowest ones. 
It should be a good reason that some representations 
are not used. Missing representations may indicate that the model  
 is at least incomplete, if not wrong: something (additional symmetries, 
principles, new physics) should be added 
to explain their absence. Searches for reason of missing reps. should 
be important element of model building.\\

{\it II. Large Discrete parameter space.} 
Model  building consists of selection of  

(i) level $N$ which determines type of finite discrete  group,     

(ii) representations of fermions,  

(iii) transformation of fermion bi-linears,  

(iv) weights of modular forms,  

(v) types of modular forms with given transformation properties in the case of high weights.    

\noindent
At each step there are multiple possibilities.   
Huge number of possible charge prescriptions form huge discrete parameter space. 
In model building the algorithm of scanning of the possibilities 
in particular,  selection of charge prescriptions 
can be elaborated which lead to agreement with observations. 
In fact, principles that select the  point (few points) in this space were never discussed.  

\section{On model building. Examples}

Minimal modular symmetry models which do not have problems with missing representations and 
big discrete parameter space contradict data.  
The model \cite{feruccio} has $A_4$ symmetry and very logical assignment of flavor charges 
(see the example 3 of  \cite{feruccio} without flavons).  
Fermions, both EW doublets and RH neutrinos,  form  the $A_4$ triplets, while RH charged leptons 
fill in the singlet representations $1, 1', 1''$.  Weights of all matter fields 
(apart from Higgses) equal 1.  Higgses have zero weights. The Yukawas form  triplets with weight 2.
Free parameters are tunned to obtain correct neutrino masses, but 
in the $\chi^2$ minimum, which corresponds to $\langle \tau \rangle = 0.008 + 0.98 i$,  
the predicted mixing angles are by many sigmas  away from the observed values. 

Scanning larger range of values of $\tau$ \cite{Ding:2019zxk} 
it was found later that value  
$\langle \tau \rangle = 0.04 + 2.23 i$ leads to  very agreement with data: $\sin^2 \theta_{12} = 0.310$, 
$\sin^2\theta_{13} = 0.0224$,
$\sin^2 \theta_{23} = 0.58$ (see model D10 in \cite{Ding:2019zxk}). 
However, in this $\chi^2-$ minimum the mass spectrum of neutrinos is quasi-degenerate
with the sum of masses $\sum_i  m_i = 0.297$ eV,  which
contradicts the upper cosmological bound. 
The smaller sum, $\sum_i  m_i = 0.178$ eV, has been found at $\langle \tau \rangle = 0.01 + 1.80 i$ 
in the recent fit which uses the updated neutrino data \cite{marrone}. 
Still this is substantially above the cosmological bound.
Furthermore,  a number of questions is left: why RH neutrinos are in triplet 
while the RH components of charged leptons singlets? This prevents extension to the LR symmetry. 
Inclusion of quarks and GUT construction are problematic. 
The smallness of coupling constants,  $g_i \sim 10^{-4}$, is not explained.  
In any case the number of fit parameters equals number of oscillation
observables. Some freedom of the charge (weight) prescription exists.

With this  one can proceed in two different ways: (i) modify (complicate) symmetry in a way that 
allows to introduce more free parameters to fit the data, (ii)  keep minimal symmetry, but 
add some new physics not related to this symmetry. 
Recall that in the case of traditional symmetry approach the model building  
traces  the symmetry of mass matrices as much as possible.
In contrast, the modular symmetry approach follows the ``direct'' procedure when symmetry 
of the Lagrangian is determined,  charge assignment 
is performed and then phenomenological consequences are obtained. 
Disagreement  with data is direct  consequence of symmetry. 
{\it At this point the fight against the symmetry starts.} 
To achieve agreement with data more free continuous parameters 
are introduced to essentially wash out traces of symmetry.  
For this  complicated symmetry is selected with many different representations.  
Thus, creating huge discrete parameter space. 
This  parameter space acts essentially as continuous parameters. \\

To illustrate  general points described above 
we discuss two  examples of models. 

\begin{figure}
\begin{center}
\includegraphics[width=0.85\linewidth]{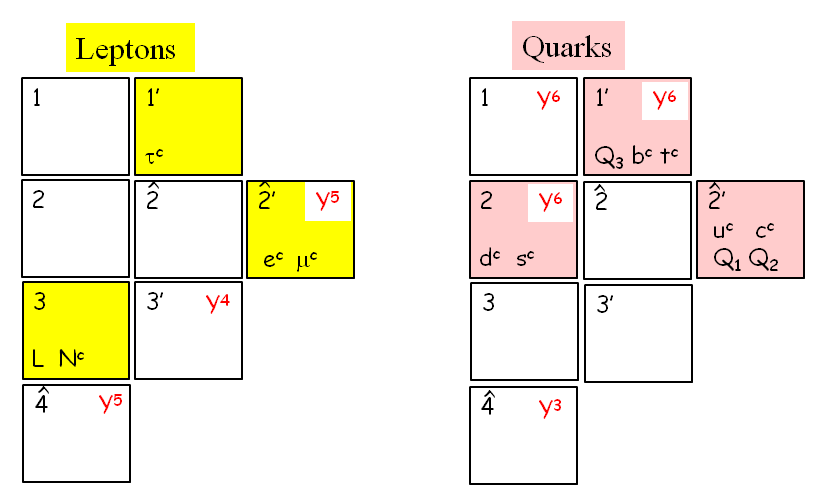}
\end{center}
\caption{
The multiplets of symmetry group. Colored fields are representations 
which filled in by particles of the model.
}
\label{fig:di}
\end{figure}

1. Models with binary octahedral symmetry \cite{ding}, \cite{marrone}. 
In Fig. \ref{fig:di} we show representations of the symmetry group and assignments 
for fermionic fields and modular forms. 
The complicated symmetry  leads to the problem of missing representations: 
Particles in the rep.  $\hat{4}, 3, \hat{2}$ 
are absent. 
Furtheremore, assignments for  leptons and  quarks are different.  
Therefore further unification and GUT  are not possible at least in a simple and economic version.

\begin{table}
\begin{center}
\begin{tabular}{|l|l|l|l|}
\hline
particles & Multiplets & representations      & $\#$ possibilities \\
\hline
L                    & T     &   3, $3'$   &   2  \\                
\hline
\, \,                & T     &   3, $3'$   &  2\\
$e^c, \mu^c, \tau^c$ & D + S & $(2, \hat{2} , \hat{2}' ) \times ( 1, 1')$   &  $6 \times3 = 18$\\
\, \,     &S + S + S & $(1 + 1')^3 $   & 8  \\
\hline
\end{tabular}
\end{center}
\label{tab1}
\caption{Possible symmetry assignment in the case of binary octahedral group}
\end{table}

Let us estimate the  discrete parameter space of the model
(see the Table 1). In the lepton sector  
even if we  assign $L$ to the  triplet, a   
large degree of  freedom of charge assignments still remain as indicated in the Table 1.   
Total number of possibilities for RH charged leptons equals  $(2 + 18 + 8) = 28$.   
The same number of assignments, 28,  is for $N_1^c$,  $N_2^c$,  $N_3^c$. 
Consequently, the total number  $28 \times 28 \times 2 = 1568$  
possibilities of discrete symmetry  assignment exists. 
If also the EW doublets $L$ are allowed to be in various representation of the group 
and not only in triplets,  we will get $(28)^3  = 21952$ possible 
charge assignments for known fermions.  
Additional possibilities are related to assignment of weights.    
Invariance can be achieved by selection of $Y(\tau)$ and weights 
of RH components of leptons for a given weight of $L$. \\

The superpotential  of the model reads 
\begin{eqnarray}
W & = & g_1^E (E_D^c L)_{\hat{2'}} Y_{\hat{2'}}^{(5)}H_d  + g_2^E (E_D^c L)_{\hat{4}} 
Y_{\hat{4}}^{(5)} H_d   +  g_3^E (\tau^c L)_{3'}  Y_{3'}^{(4)} H_d 
\nonumber\\ 
& + & g (N^c L)_1 H_u  
 + \Lambda (N^c N^c )_2 Y_2^{(2)} H_d , 
\label{eq:superpot}
\end{eqnarray}
where  $E_D^c \equiv (e^c, \mu^c)$.  The group structures of fermion bilinears 
in the first  line  of Eq. (\ref{eq:superpot})   
are $(2'\times 3 = 2' + 4)$ as well as $(1'\times 3 = 3')$, and in the second line:  
$(3\times3 = 1 + 2 + 3  +3') $. 

Let us see how the "successful" fit of the data is realized: 
Observables include  3 mixing angles, 3 charged lepton masses, three neutrino masses. 
Four continuous fit parameters  
($g^E_1$,  $g^E_2$, $g^E_3$,  $v_d$), 
were used to reproduce 4 observables
$m_e$, $m_\mu$, $m_\tau$ and  $\theta_{13}$. 
Three continuous parameters $(g^2 v_u^2/\Lambda)$, 
$\tau^R$ , $\tau^I$, 
where $v_d$ is the Higgs VEV,  allow to get  
correct values of three neutrino masses $m_1$, $m_2$ and $m_3$ or equivalently, 
$m_1$, $\Delta m_{21}^2$ and  $\Delta m_{31}^2$.   
Discrete parameters determine values 
of continuous parameters  
$\theta_{12}$ and  $\theta_{23}$.  Important element 
of the game is that the discrete parameters can fix values of continuous parameters. 
Indeed, the modular forms in the neutrino sector are selected in such a way that  
state 1 decouples while the $\mu \mu$- and $\tau\tau$-  elements of the mass matrix 
are equal which leads to  maximal 2-3 mixing.   
Contribution from the charge lepton sector  is small. 

The best fit values of couplings equal 
$g^E_1 = 1.4 \cdot 10^{-5}$,   
$g^E_2 = 0.716 g^E_1 = 1.0 \cdot 10^{-5}$,  
$g^E_3 = 87.7 g^E_1 = 1.23 \cdot 10^{-3}$ 
for $v_d = 5$ GeV and  
the origin of this hierarchy of the couplings is not explained. \\

\begin{figure}
\begin{center}
\includegraphics[width=0.4\linewidth]{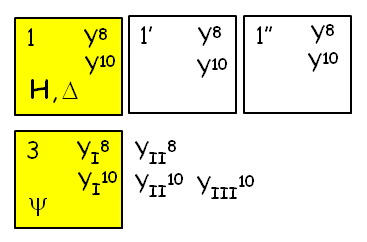}
\end{center}
\caption{
$SO(10)$ with modular flavor symmetry.
}
\label{fig:w}
\end{figure}

As an example of GUT,  let us consider the  SO(10) model 
with modular flavor $A_4$ \cite{king}. 
The minimal SO(10) model (model 2 of \cite{king}) has  $10 (H) + \overline{126} (\bar{\Delta})$  Higgses,  
and three 16 plets $\psi$.  
The symmetry assignment is shown in Fig. \ref{fig:w}. 
Here the way to introduce enough free parameters is to assign  
large values of weights to the modular forms $Y$:
from  possible values of discrete parameters $k_Y  = 0, 2, 4, 6, 8, 10, ...$  
$k_Y = 8$ and  $k_Y = 10$ are selected. 
Invariance ensured by corresponding choice of $k_H$ for scalars.  (For  $k_Y^{(10)}$ and $k_Y^{(6)}$ 
there are 36 possibilities, but among them only 3 are viable and the fit is not perfect.)
For large weight, $Y$ can be in any $A_4$   representation, furthermore   
Rep. 3 can be in three versions related to  permutation of  columns.  
This allows one to introduce many interaction terms with arbitrary couplings:
\begin{eqnarray}
W & = & \alpha_1 Y_1^{(10)}(\psi \psi H)_1 
+ \alpha_2 Y_{1'}^{(10)}(\psi \psi H)_{1''}   
 +  \left[\alpha_3 Y_{3I}^{(10)} 
+ \alpha_4 Y_{3II}^{(10)} 
+ \alpha_5 Y_{3III}^{(10)}\right](\psi\psi H)_3
\nonumber\\  
 & + & \gamma_1 Y_1^{(8)} (\psi \psi \bar{\Delta})_1 
+ \gamma_2 Y_{1'}^{(8)}(\psi \psi \bar{\Delta})_{1''} 
+ \gamma_3 Y_{1''}^{(8)}(\psi \psi \bar{\Delta})_{1'}
+ \left[\gamma_4 Y_{3I}^{(8)} + \gamma_5 Y_{3II}^{(8)}
\right] (\psi \psi H)_{3}.  
\label{eq:wforso10}
\end{eqnarray}
Here $k_\psi = 5$, $k_{10} = 0$ and $k_{126} = -2$.  
Couplings  $\alpha_i$, $\gamma_j$,  and VEVs of  $\tau$  
compose 27 real parameters. 
On the other hand there are 27 observables: masses and  mixings of quarks  and leptons. 
The  best fit values of parameters spread by 2.5 orders of magnitude. 
Predictions are given for unknown yet quantities: the 3 CP-phases 
in lepton sector, and the absolute scale of neutrino mass. 
Here the question with missing reps. still exists: why the higher weights 8 and 10 are used only? 
Why $Y$ with lower weights are absent?

\section{Further complications and ``pollution"}

It could be a long way from fundamental level at high 
mass-energy scales  down to observables at low scales. 

1. At the fundamental level,  where symmetry is realized,  the primary flavor structures are formed.  
Below this  scale the symmetry is broken. Decoupling of heavy degrees of freedom,  
as, e.g.,  in the Seesaw mechanism,  modifies flavor structure of the rest of model  at low energies. 
So, some flavor features at low energies can be a result of decoupling 
of heavy degrees of freedom.

2. The RGE effects at all the scales can further 
modify the flavor structure.

3. ``Pollution" is plausible. More elements should be included in the flavor physics:  dark matter, axions ... 
multiple contributions. As a result we have for the mass matrices the sum  
$$
M = M_1 + M_2 +  ... 
$$
Individual terms $M_i$  may have different specific symmetries but sum does not look symmetric at all  
and it is difficult to recognize symmetries of individual terms. 
In addition, contributions from physics at very low energy scales is  possible. 

``Sterile pollution" is due to presence of new fermionic degrees of freedom - singlets of SM. 
Let us consider single sterile neutrino with mass  $m_4  > 1$  eV 
that  mixes with active neutrinos, as motivated by 
several anomalies in neutrino oscillation experiments. 
The mass matrix of 4 neutrinos in the flavor basis $(\nu_e, \nu_\mu, \nu_\tau, \nu_S)$ reads 
\begin{equation}
m = 
\left(\begin{matrix}
m_{ee} & m_{e\mu} & m_{e\tau}& m_{eS}\\
... & m_{\mu \mu} & m_{\mu \tau} & m_{\mu S} \\
... & ... &  m_{\tau \tau} & m_{\tau S}\\
... & ... & ... &  m_{SS}
\end{matrix}
\right)
\label{eq:massmatr}
\end{equation}
with  $m_{ss} \approx m_4$. The active-sterile mixing  parameters  required by the data equal  
$$
\sin \theta_{eS} \approx m_{eS}/m_{SS}  \approx 0.02 - 0.1,   
$$
{\it etc.}  After decoupling of the  4th neutrino 
({\it a la} see-saw) we have   
$$
m_\nu =  m_a  + m^I,  
$$
where $m_a$ is the $3\times3$ submatrix of 
active neutrinos in (\ref{eq:massmatr}), and 
$m^I$ is the induced mass matrix with elements   
\begin{equation} 
m^I_{\alpha \beta} = 
- \sin \theta_{\alpha s}  
\sin \theta_{\beta s}  m_4,   \, \, \,  \sin \theta_{\alpha S}\approx m_{\alpha S}/ m_{SS}.  
\label{eq:ind-el}
\end{equation}
The upper bounds on elements of the induced mass matrix (\ref{eq:ind-el}) 
imposed by the oscillation data are shown in Fig. 4  \cite{pjs}.  
According to the figure, for $m_4  > 2$ eV the matrix 
$m^I$ can reproduce values of all masses  
$m_{\alpha \beta}$ for both mass orderings of light neutrinos. For $m_4  = 1$ eV  
all but the $ee-$element (in the case of IO) can be obtained from  $m^I$.
In the case of IO the nearly democratic mass matrix  
with $m \sim 2 \cdot 10^{-2}$ eV is possible. The induced mass 
$m^I$ can explain difference of quark and lepton mixings if  $m_a \approx m_{quark}$.\\
\begin{figure}
\begin{center}
\includegraphics[width=1.0\linewidth]{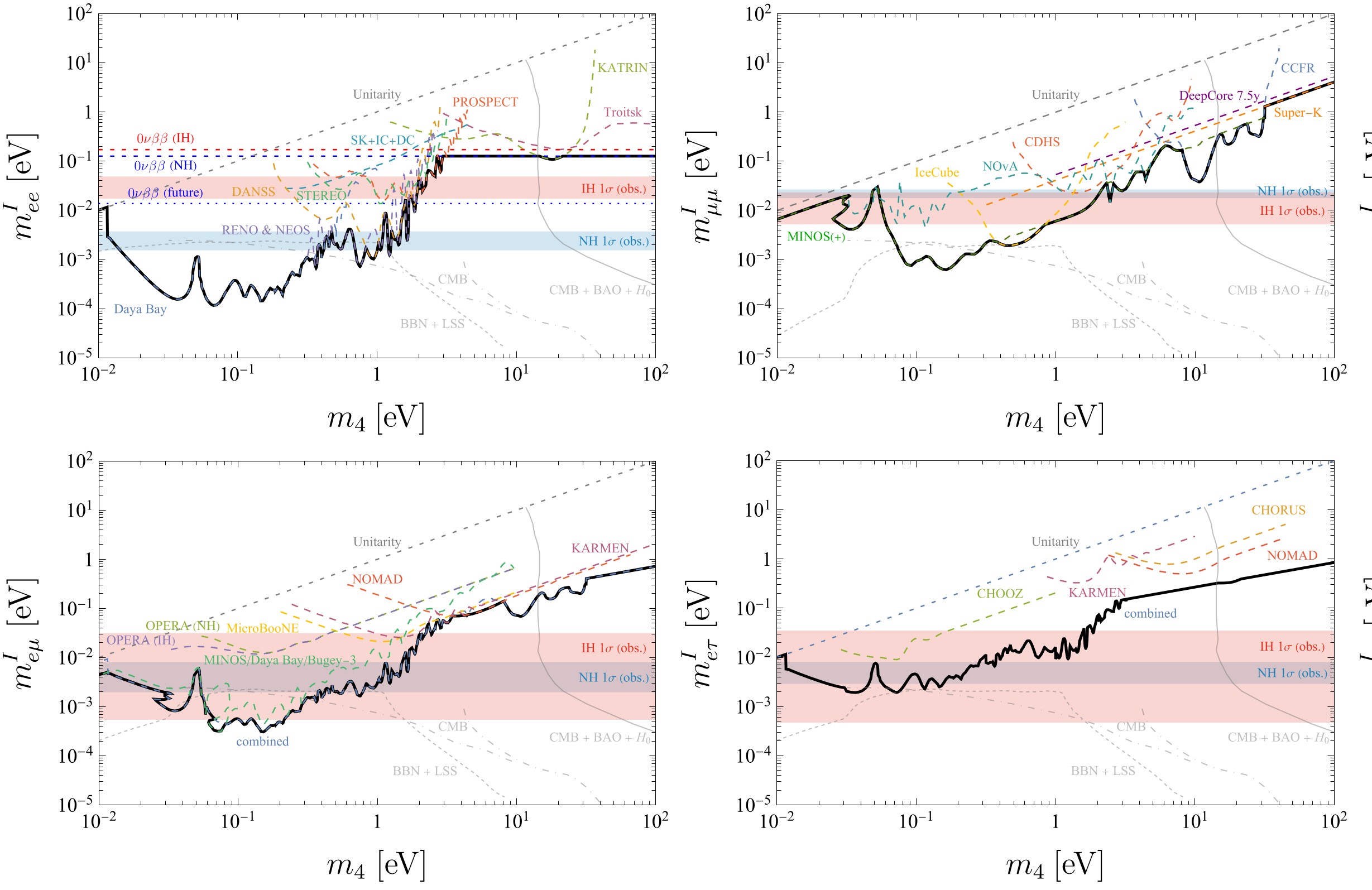}\\
\includegraphics[width=0.5\linewidth]{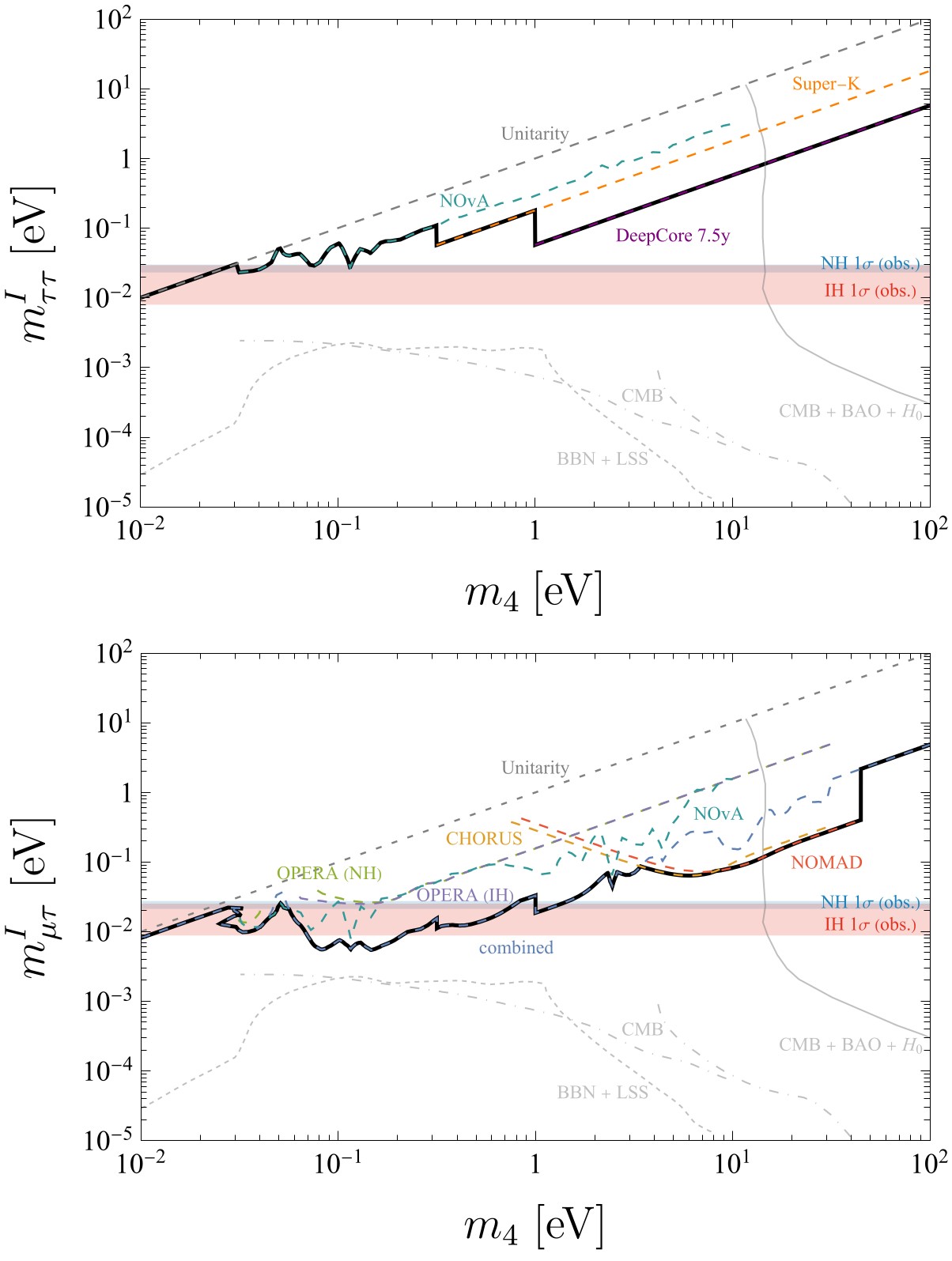}\
\caption{
Bounds on induced elements of neutrino mass matrix  from the oscillation 
experiments as function of 4th neutrino mass. 
Blue (red) bands corresponds to the allowed ranges 
of elements of mass matrix in the case of normal (inverted) mass ordering.
}
\end{center}
\label{fig:induced}
\end{figure}

Let us consider few special  cases with discrete symmetries. 
The active-sterile mixing matrix elements
$$
(m_{e S} , m_{\mu S}, m_{\tau S})  
\approx m_0 (1, 1, 1) 
$$
can be generated by the flavor-blind  $\nu_S  - \nu_a$ interactions. 
In this case  $m^I$  has the ``democratic" 
form, while $m_a$ being  subdominant gives required  corrections. 
This can be realized as a consequence of $S_3$ symmetry for active neutrinos. 

The vector of mixing mass terms 
$$
(m_{e S} , m_{\mu S}, m_{\tau S})  
\approx m_0 (0, 1, 1) 
$$
generates the 
dominant $\mu \tau$-block with small determinant.  This gives 
maximal  2-3 mixing and enhances  1-2 mixing. 
The underlying symmetry could be $S_2$.  The structure
$$
(m_{e S} , m_{\mu S}, m_{\tau S})  
\approx m_0 (0, 0, 1) 
$$
allows to suppress $m_{\tau \tau}$ element of mass matrix,  and consequently,  
enhance 2-3 mixing.   $Z_2$ symmetry can  be behind. \\


Flavor structures can be affected or even formed by interactions with Dark sector 
\cite{Ludl:2015tha} \cite{Smirnov:2018luj} 
(see Fig. 5).  The lepton mixing matrix may have the following form

\begin{figure}
\begin{center}
\includegraphics[width=0.6\linewidth]{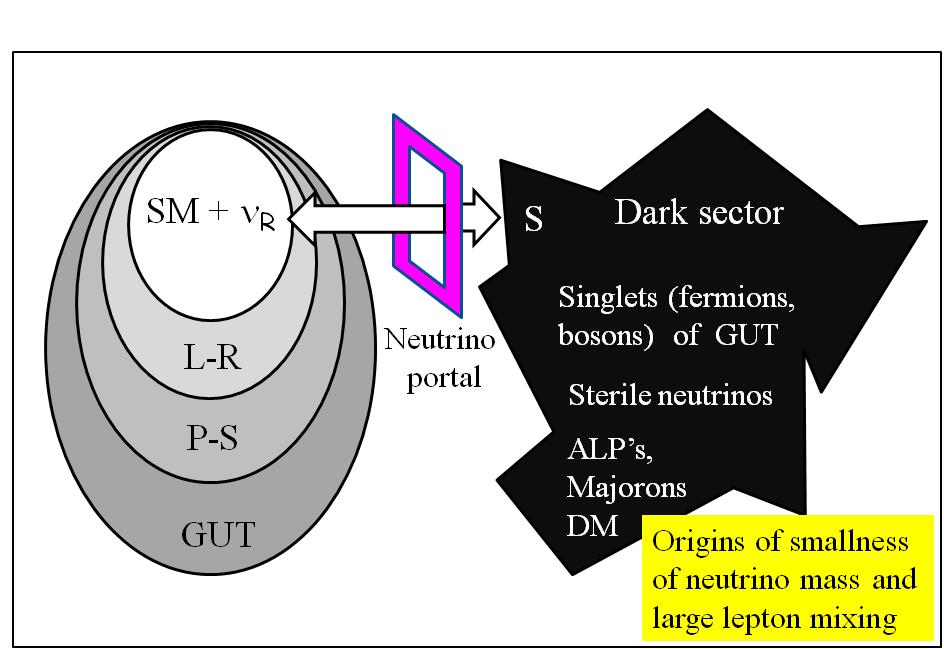}
\caption{
Set up with hidden sector.
}
\end{center}
\label{fig:set}
\end{figure}
\begin{equation}
U_{PMNS} = V_{CKM}^\dagger U_X,  
\label{eq:qlcompl}
\end{equation}
where the CKM-like contribution, $V_{CKM}$, originates from common sector for quarks  and leptons. 
This implies   $m_l \sim m_d$,  $m^D_\nu \sim m_u$,  the 
$q - l$ unification,  GUT.    
This sector  produces the CKM structures, hierarchy  
of masses and mixings,  Froggatt-Nielsen  type   relations 
between masses and mixing.  The matrix $U_X$ emerges from the dark sector coupled  to active neutrinos 
via the RH neutrino portal. It is responsible  for 
large neutrino mixing and smallness of neutrino mass. 
Furthermore, it  may have special symmetries  which lead  
to the BM (bi-maximal) or TBM mixing. In fact,  it is easy to realize symmetries responsible for BM or TBM 
in the  hidden sector and then propagate their effects to the visible sector via the neutrino portal 
with certain basis fixing symmetry.  
For $U_X= U_{TBM}$ or $U_{BM}$  Eq. (\ref{eq:qlcompl}) leads to the relation  
$$
\sin^2\theta_{13} \approx \sin^2\theta_{23}\sin^2\theta_{C}   
$$
which is in a good agreement with measurement. 
Here $\theta_{C}$ is the Cabibbo angle.  So,  the 1-3 mixing is the result of TBM  in the Dark sector 
and the $q-l$ unification. 
There is the  GUT- Planck scales realization of this scenario 
\cite{Smirnov:2018luj}. The simplest basis fixing symmetry is   $G_{basis} = Z_2 \times Z_2$. 

\section{Conclusions and Outlook}

Relevance of discrete symmetries for explanation of the observed flavor structures is still an open question. 
Apparent randomness in data, lack of simple description of the data in terms of small number 
of parameters give little hope that simple and exact explanation exists. 
The main flavor features: mass hierarchy of the charged leptons and large lepton mixing are not explained. 

In models with modular symmetries the flavon dependent Yukawa couplings $ \langle \phi \rangle/\Lambda $ are substituted  
by Modular forms which depend on sigle (in the simplest case) moduli. 
They act in the same way as flavon multiplets with specifically selected VEV's. 
Although models with modular symmetries are expected to be more restrictive, 
the agreement with data require introduction of addition freedom which reduces 
predictivity to the same level as flavon models. 

The simplest modular models  do not agree with data.  
Complications of symmetries and models are needed to introduce more free  
parameters (continuous and discrete).  This complication leads to two problems: 
\begin{itemize}

\item
 missing representations,  and 

\item 
large discrete parameter space. 

\end{itemize}
The problems should be resolved in complete and consistent 
models.

  The observed flavor features follow from fitting of  free parameters 
and not  from structure of  formalism. 
In the  proposed models (based on  both traditional and modular symmetries) matching with data 
is interplay of large number of continuous parameters and {\it ad hoc} selection 
of specific points in huge discrete parameter space related to group and charge assignments.  
Structure of models (mathematical formalism) 
does not map well onto set of observables.
Model building is reduced essentially to the fight 
against the symmetry. 

Probably  two or more different and  independent  contributions to the neutrino mass matrix exist. 
One possibility is the contribution from mixing of usual neutrinos with sterile neutrinos. 
Another one is from the Dark sector (with discrete symmetries) via the neutrino portal. \\ 

\newpage

\noindent
{\bf Note added.}\\

In \cite{Baretz:2025zsv} that appeared after the conference 
the Autonomus Model Builder (AMBer) has been elaborated which allows one to scan 
possibilities within certain flavor symmetry framework. That is, select symmetry groups, 
particle content, group representation assignment and in this way construct 
viable models with minimal number of free parameters.  
This is awaited  and really desirable step which 
simplifies endless and in a sense trivial work of "handmade" model building.  
AMBer is very useful for those who want to proceed with 
already known framework. But probably searches of something qualitatively new (concepts, principles) is needed 
which AMBer does not yet include.

\section*{Acknowledgments}
The author is grateful to A. Trautner for numerous discussions. 
I would like to thank Feruccio Feruglio and Antonio Marrone for discussions of relevance of the modular 
symmetry approach during the workshop and email exchange after the workshop.

\end{document}